\documentclass{article}

\usepackage[nonatbib, final]{ml4mols_2020}

\usepackage[utf8]{inputenc} 
\usepackage[T1]{fontenc}    
\usepackage{hyperref}       
\usepackage{url}            
\usepackage{booktabs}       
\usepackage{amsfonts}       
\usepackage{nicefrac}       
\usepackage{microtype}      

\usepackage{hyperref}
\hypersetup{
    colorlinks=true,
    linkcolor=blue,
    filecolor=magenta,      
    urlcolor=cyan,
}
\usepackage{graphicx}
\usepackage{bm}
\usepackage[ruled,vlined]{algorithm2e}
\usepackage{pifont}

\usepackage{amsfonts}
\usepackage{adjustbox}

\newcommand{\argmax}{\arg\!\max}
\DeclareUnicodeCharacter{0301}{\'{e}}

\usepackage{wrapfig}

\title{Hyperbolic Molecular Representation Learning for Drug Repositioning}

\author{%
  Ke Yu \\
  University of Pittsburgh\\
  \texttt{yu.ke@pitt.edu} \\
  \And
  Shyam Visweswaran \\
  University of Pittsburgh\\
  \texttt{shv3@pitt.edu} \\
   \AND
  Kayhan Batmanghelich \\
  University of Pittsburgh\\
  \texttt{kayhan@pitt.edu} \\
}

\begin{document}

\maketitle

\begin{abstract}
Learning accurate drug representations is essential for task such as computational drug repositioning. A drug hierarchy is a valuable source that encodes knowledge of relations among drugs in a tree-like structure where drugs that act on the same organs, treat the same disease, or bind to the same biological target are grouped together. However, its utility in learning drug representations has not yet been explored, and currently described drug representations cannot place novel molecules in a drug hierarchy. Here, we develop a semi-supervised drug embedding that incorporates two sources of information: (1) underlying chemical grammar that is inferred from chemical structures of drugs and  drug-like molecules (unsupervised), and (2) hierarchical relations that are encoded in an expert-crafted hierarchy of approved drugs (supervised). We use the Variational Auto-Encoder (VAE) framework to encode the chemical structures of molecules and use the drug-drug similarity information obtained from the hierarchy to induce the clustering of drugs in hyperbolic space. The hyperbolic space is amenable for encoding hierarchical relations. Our qualitative results support that the learned drug embedding can induce the hierarchical relations among drugs. We demonstrate that the learned drug embedding can be used for drug repositioning.
\end{abstract}

\section{Introduction}
The study of drug representation provides the foundation for computational drug repositioning. Drug repositioning, the process of finding new uses for existing drugs, is one strategy to shorten the time and reduce the cost of drug development~\cite{nosengo2016new}. Computational methods of drug repositioning typically aim to identify shared mechanism of actions among drugs that imply that the drugs may also share therapeutic applications~\cite{pushpakom2019drug}. However, such methods are limited when prior knowledge of drugs may be scarce or not available; for example, drugs that are in the experimental phase or have failed clinical trials. Therefore, it is appealing to map the chemical structure of a molecule to its pharmacological behavior.

A drug hierarchy encodes a broad spectrum of known drug relations. For example, a widely used drug hierarchy, Anatomical Therapeutic Chemical Classification System (ATC), groups drugs that are similar in terms of their mechanism of action and therapeutic, pharmacological and chemical characteristics. But its utility in learning drug representation has not yet been explored.

\begin{figure*}[t]
    \centering
    \includegraphics[width=1.0\textwidth]{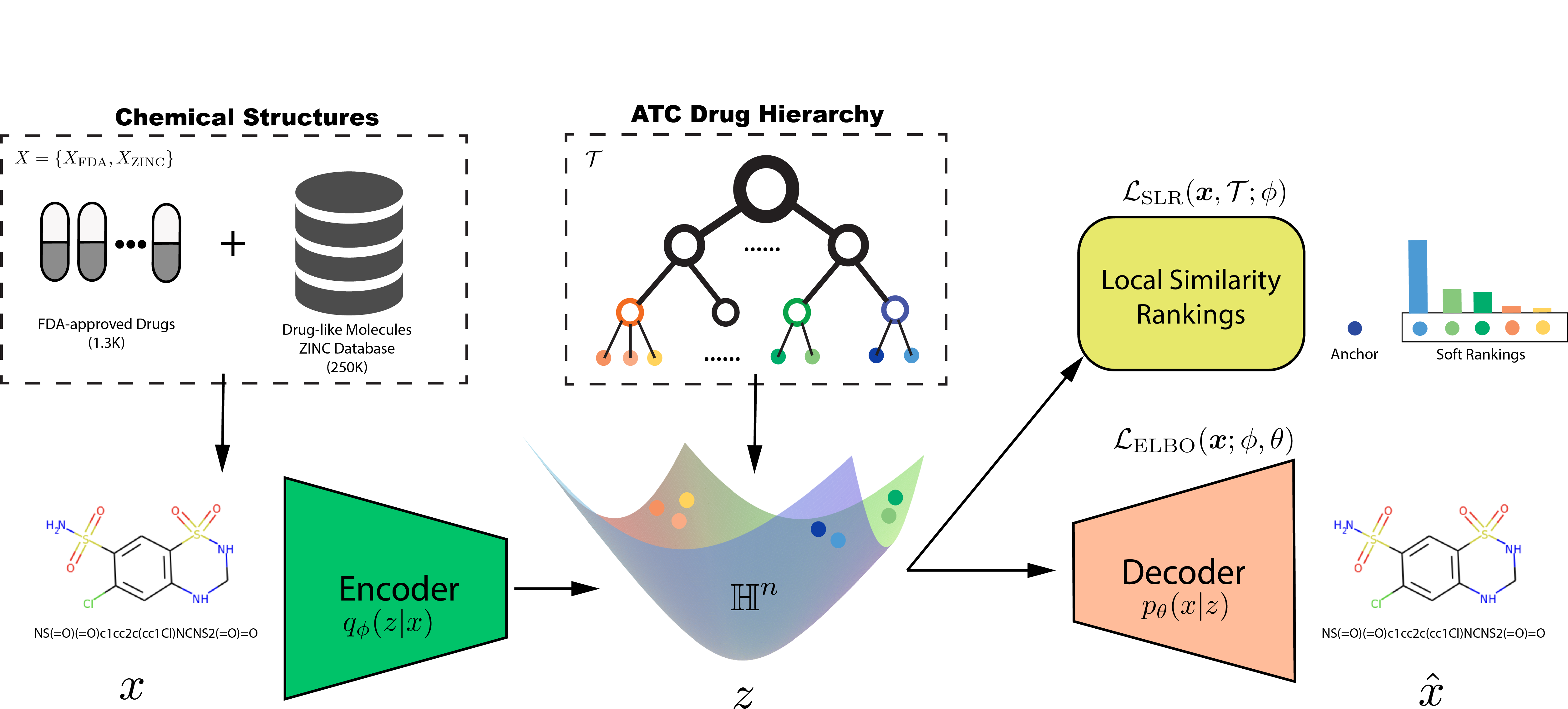}
    \caption{Schematic diagram of the proposed drug embedding method. Our semi-supervised learning approach integrates the chemical structures of a small number of FDA-approved drug molecules ($X_\mathrm{FDA}$) and a larger number of drug-like molecules ($X_\mathrm{ZINC}$) drawn from the ZINC database. We use VAE to encode molecules in hyperbolic space $\mathbb{H}^n$, and enforce the ATC drug hierarchy by preserving local similarity rankings of drugs. The symbols $\bm{x}$, $\bm{z}$, $\hat{\bm{x}}$ denote a molecule represented by its SMILES string, its embedding and its reconstruction; $q_\phi(\bm{z}|\bm{x}), p_\theta(\bm{x}|\bm{z})$ denote the encoder network and the decoder network respectively; $\mathcal{L}_\mathrm{ELBO}(\bm{x};\phi,\theta)$, $\mathcal{L}_\mathrm{SLR}(\bm{x},\mathcal{T};\phi)$ denote the objective functions for the VAE and the local similarity rankings.}
    \label{fig:1}
\end{figure*}

Here, we develop a drug embedding that integrates the chemical structures of drugs and drug-like molecules with a drug hierarchy such that the similarity between pairs of drugs is informed both by the structure and groupings in the hierarchy (Figure~\ref{fig:1}). To learn the underlying grammar of chemical structures, we leverage a data set of drugs (about 1.3K) that are approved by the Food and Drug Administration (FDA) and a larger data set of drug-like molecules (about 250K) and use the simplified molecular-input line-entry system (SMILES)~\cite{weininger1988smiles} structure representation. We obtain drug similarity relationships from the ATC drug hierarchy. We use the hyperbolic space for the embedding since it is amenable for learning continuous concept hierarchies~\cite{nickel2017poincare, mathieu2019continuous,de2018representation, monath2019gradient}.

\section{Method}

\textbf{Learning chemical grammar using VAE}
We use a Variational Auto-Encoder (VAE) to encode the chemical structure of molecules. More specifically, we model a molecule as a random variable generated by encoding a SMILES string through a encoder $q_\phi(\bm{z}|\bm{x})$ into a code ($\bm{z}$), which is then decoded back to a reconstruction of the input by passing through a decoder $p_{\theta}(\bm{x}| \bm{z})$. In the classic VAE, the prior $p(\bm{z})$ is the standard normal distribution, the encoder $q_\phi(\bm{z}| \bm{x})$ is modeled by a Gaussian distribution $\mathcal{N}(\bm{z}|\bm{\mu}_\phi, \Sigma_\phi)$. To extend VAE from a flat Euclidean space to a curved manifold, the Gaussian distribution needs to be extended to the hyperbolic space.

In this study, we adopt the so-called wrapped normal distribution proposed by Nagano et al., 2019~\cite{nagano2019wrapped}, which we denote by $\mathcal{N}_{\mathbb{H}}^{\mathrm{W}}(\bm{z}|\bm{\mu}, \Sigma)$, where $\mathbb{H}^n$ is the Lorentz model of a $n$-dimensional hyperbolic space, $\bm{z} \in \mathbb{H}^n$, and $\bm{\mu}$ is the hyperbolic mean. The reparameterization trick in the hyperbolic VAE can be viewed as the composition of two operations $\mathrm{exp}_{\bm{\mu}}(\mathrm{PT}_{\bm{\mu}_0 \rightarrow \bm{\mu}} (\bm{u}))$, where $\mathrm{PT}_{\bm{\mu}_0 \rightarrow \bm{\mu}} (\bm{u})$ is the so-called \textit{parallel transport}, $\mathrm{exp}_{\bm{\mu}}$ is the so-called \textit{exponential map}, and $\bm{u} \sim \mathcal{N}(\bm{0},\Sigma)$ is a random vector sampled from normal distribution. The inside operation $\mathrm{PT}_{\bm{\mu}_0 \rightarrow \bm{\mu}} (\bm{u})$ shifts the tangent space, a linear approximation of the manifold around a point, from $\bm{\mu}_0$ to $\bm{\mu}$ analogous to the addition operation in the classic reparameterization trick. The $\mathrm{exp}_{\bm{\mu}}$ projects the shifted vector to the manifold. Note that, in the Lorentz model, both the \textit{parallel transport} and the \textit{exponential map} have analytical forms, 
and can be differentiated with respect to the hyperbolic mean $\bm{\mu}$ of the wrapped normal distribution $\mathcal{N}_{\mathbb{H}}^{\mathrm{W}}(\bm{z}|\bm{\mu}, \Sigma)$. We compute the KL-divergence following the derivation in~\cite{nagano2019wrapped}.

\textbf{Integrating hierarchical knowledge} The hyperbolic VAE learns an embedding for codes that are amenable to hierarchical representation. However, it only models $\bm{x}$ (the SMILES string of the drug), and it does not enforce our prior knowledge about drug hierarchy which defines similarity or dissimilarity between drugs at various levels. Inspired by concept embedding in hyperbolic space~\cite{2018arXiv180603417N}, we incorporate the ATC hierarchy $\mathcal{T}$ in our model by using pairwise similarity between drugs.
Let $t_{i,j}$ denote the path-length between two drugs, $\bm{x}_{i}$ and  $\bm{x}_{j}$ in $\mathcal{T}$, and let $\mathcal{D}(i,j)=\{k: t_{i,j}<t_{i,k}\} \cup \{j\}$ denote the set of drugs with path-lengths equal to or greater than $t_{i,j}$. We define the soft local ranking with respect to the anchor drug $\bm{x}_i$ as:
\begin{equation}
\label{eq:slr}
    p(\bm{x}_i, \bm{x}_j;\phi) = \frac{\mathrm{exp}(-d_{\ell}(\bm{\mu}_i, \bm{\mu}_j))}{\sum_{k \in \mathcal{D}(i,j)} \mathrm{exp}(-d_{\ell}(\bm{\mu}_i, \bm{\mu}_k))}
\end{equation}
where $\bm{\mu}_i$ is the hyperbolic mean of $q_\phi(\bm{z}|\bm{x}_i) = \mathcal{N}_{\mathbb{H}}^{\mathrm{W}}(\bm{z}|\bm{\mu}_i, \Sigma_i)$ and $d_{\ell}(\bm{\mu}_i, \bm{\mu}_j)$ is the hyperbolic distance between $\bm{\mu}_i$ and $\bm{\mu}_j$. The likelihood function of the soft local rankings is given by $\mathcal{L}_{\mathrm{SLR}}(\bm{x}_i, \mathcal{T}; \phi) = \sum_j \mathrm{log}\ p(\bm{x}_i, \bm{x}_j;\phi)$. 

Note that the global hierarchy of $\mathcal{T}$ is decomposed into local rankings denoted by $\mathcal{D}(i,j)=\{k: t_{i,j}<t_{i,k}\} \cup \{j\}$. To train our model, we need to effectively sample $\mathcal{D}(i,j) \sim \mathcal{T}$, and the best sampling strategy supported by the empirical results of ablation study is as follows. For each anchor drug $\bm{x}_i$, we uniformly sample a positive example $\bm{x}_j$, such that the lowest common ancestor of $\bm{x}_i$, $\bm{x}_j$ has equal chance of being an internal node at any level, i.e., level 1, 2, 3, or 4, in the ATC tree. We then randomly sample $k$ negative examples $\bm{x}_k$ from other leaf nodes that have greater path lengths than $t_{i,j}$. 

\textbf{Optimization} We employ a semi-supervised learning approach that combines a small number of FDA-approved drugs $X_\mathrm{FDA}$ with a larger number of drug-like molecules $X_\mathrm{ZINC}$. We upsample $X_\mathrm{FDA}$ to $20\%$ in mini-batch to enhance the signal of the supervised learning task, which is maximizing the likelihood of the soft local rankings with respect to the ATC hierarchy $\mathcal{T}$. The unsupervised learning task is to maximize the variational evidence lower bound (ELBO)~\cite{jordan1999introduction} of the marginal likelihood of the chemical structures of drugs and drug-like molecules $X=\{X_\mathrm{ZINC}, X_\mathrm{FDA}\}$. We then formulate the drug embedding problem as:
\begin{equation}
\label{eq:obj}
     \argmax_{\phi, \theta} \Big(\mathcal{L}_{\mathrm{ELBO}}(\bm{x}; \phi, \theta)
      + c\cdot\mathcal{L}_{\mathrm{SLR}}(\bm{x}, \mathcal{T}; \phi) \Big)
\end{equation}
where $c=1$ when $\bm{x} \in X_\mathrm{FDA}$, $c=0$ when $\bm{x} \in X_\mathrm{ZINC}$, and $|X_\mathrm{ZINC}| \gg |X_\mathrm{FDA}|$. The first term in the objective function captures the underlying chemical grammar of molecules, and the second term enforces the relative positions of the drugs in the latent space to correspond to their relative positions in the ATC hierarchy. Parameters are estimated using mini-batch gradient descent and gradients are straightforward to compute using the hyperbolic reparameterization trick.

\section{Experimental Results}
\textbf{Datasets} We obtained SMILES strings of 1,365 FDA-approved drugs and SMILES strings of 250,000 drug-like molecules extracted at random
by~\cite{gomez2018automatic} from the ZINC~\cite{irwin2012zinc} database. We combine the 1,365 drug and the 250,000 drug-like molecules to create a single data set of chemical structures that we use in our experiments. The ATC hierarchy was created by the World Health Organization (WHO)~\cite{world2014collaborating} that leverages the location of action, therapeutic, pharmacological and chemical properties of drugs to group them hierarchically. We obtained the ATC hierarchy from the Unified Medical Language System (UMLS) Metathesaurus (version 2019AB) and mapped the FDA-approved drugs to the terminal nodes in the ATC tree that represents the active chemical substance.

\begin{figure*}
    \centering
    \includegraphics[width=0.85\textwidth]{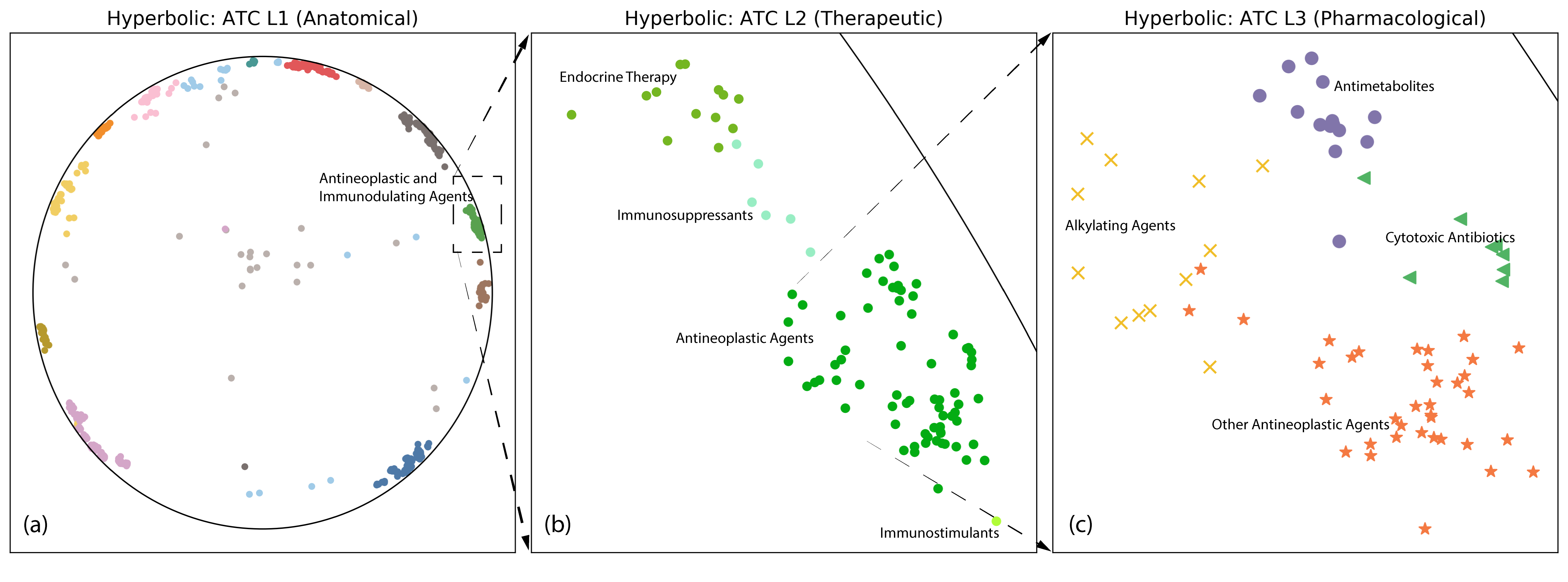}
    \caption{Visualization of hyperbolic drug embedding in two-dimensional Poincar\'e disk that shows drugs with colored symbols. In panel (a) drugs that belong to the same group at ATC level 1 are denoted by circles of the same color. Panel (b) shows drugs of one group from ATC level 1 namely, "Antineoplastic and Immunodulating Agents", and drugs that belong to the same group at ATC level 2 are denoted by circles with the same shade of green. Panel (c) shows drugs of one group from ATC level 2, namely, "Antineoplastic Agents", and drugs that belong to the same group at ATC level 3 are denoted by symbols of the same color.}
    
    \label{fig:emb_2d}
\end{figure*}

\textbf{Model visualization} We visually explore the embedding in two dimensional hyperbolic space by mapping the embedding in the Lorentz model to the Poincar\'e disk via a diffeomorphism described in~\cite{nickel2018learning}. In Figure~\ref{fig:emb_2d}(a), we observe that most of the drugs are placed near the boundary of the Poincar\'e disk and form tight clusters that correspond to the drug groups at ATC level 1. The hyperbolic embedding exhibit a clear hierarchical structure where the clusters at the boundary can be viewed as distinct substrees with the root of the tree positioned at the origin. A small number of drugs (grey circles) are scattered around the origin and denote drugs that act on the sensory organs. This group of drugs mainly consist of anti-infectives, anti-inflammatory agents, and corticosteroids, most of which act on more than one system and have multiple therapeutic uses. We hypothesize that these sensory organ drugs are placed close to the center because minimizing the local ranking loss constrains them to be concurrently close to different drug groups in the latent space.  Figure~\ref{fig:emb_2d}(b) and (c) demonstrate that embedding in hyperbolic space can effectively induce a multi-level tree and the embedding retains the hierarchical structure to the deepest levels.

\textbf{Evaluating drug repositioning} We evaluate the learned embedding for drug repositioning by deriving kNN models to discriminate between approved and unapproved drug-indication pairs in the repoDB~\cite{brown2017standard} dataset, a benchmark data set that contains information on drug repositioning. We tag each drug-indication pair with the date when the drug was first approved by the FDA. We choose 2000 as the cutoff year to split the repoDB data set into training (earlier than year 2000) and test (year 2000 and later) sets. The ratio between the size of training and test data sets is about $85\%:15\%$. For each drug $\bm{x}_i$ in the test set, we first encode it into the latent space using its SMILES string as the input, and then retrieve its $k$ nearest neighbors $\{X_{kNN}\}$ from the training set in the latent space. We apply majority voting to the retrieved drug-indication pairs in $\{X_{kNN}\}$ to predict the status of each indication associated with $\bm{x}_i$. For indications of $\bm{x}_i$ that do not exist in $\{X_{kNN}\}$, we assume that it has an equal probability of being either being successfully approved or failed to be approved, and we assign an equal vote (0.5) to each class.

\begin{figure}[h]
    \centering
    \includegraphics[width=0.7\textwidth]{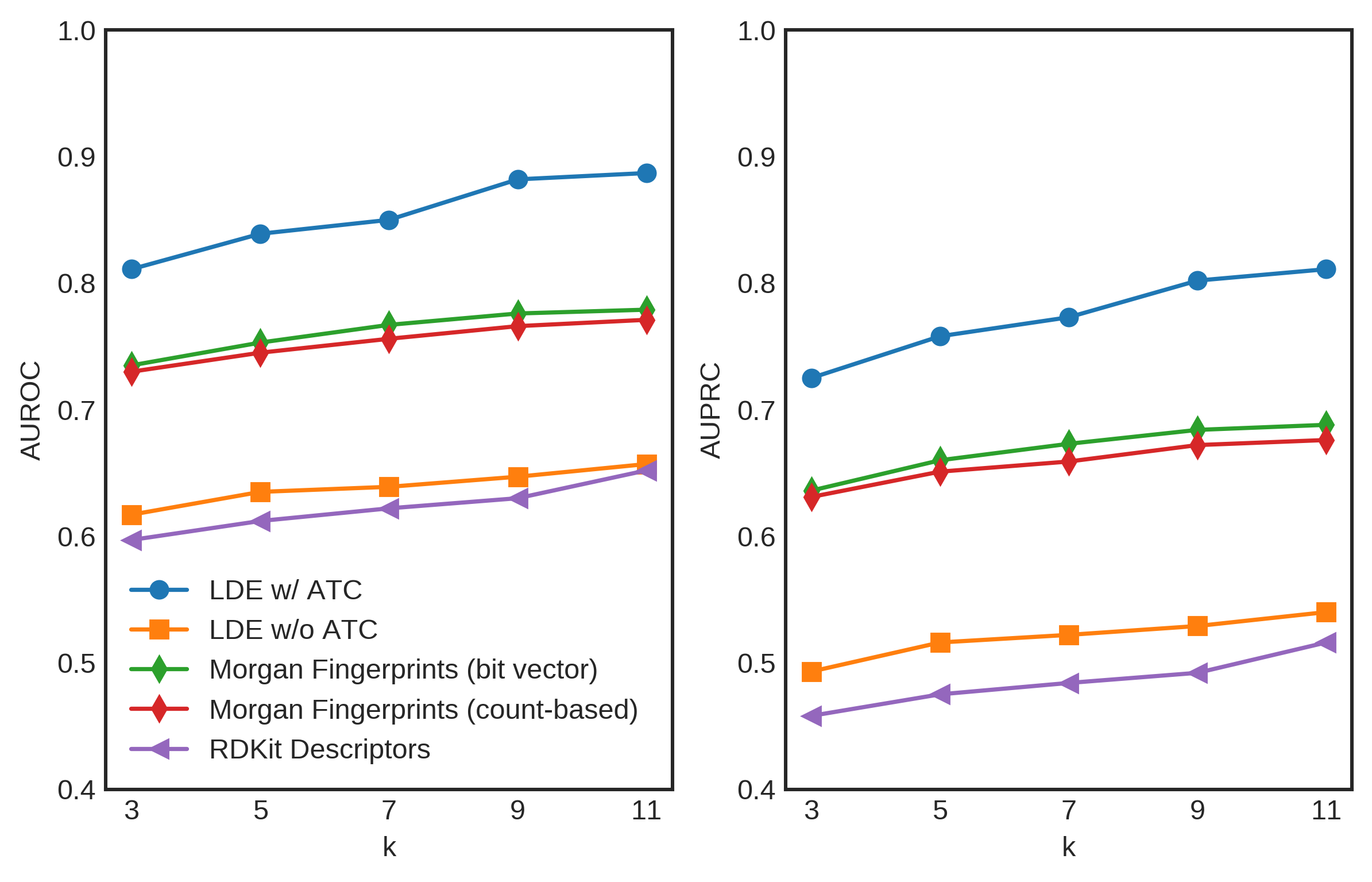}
    \caption{Comparison of representations for drug repositioning prediction using kNN (k $\in [3,5,7,9,11]$). The left panel shows AUROC scores and the right panel shows AUPRC scores.}
    \label{fig:repo_comp}
\end{figure}

Because we are not aware of any other approach developed on the repoDB dataset with the same chronological split, we compare the performance of our drug embedding, denoted as Lorentz Drug Embedding (LDE), for drug repositioning using kNN to the following baselines: (1) kNN on RDKit-calculated~\cite{landrum2006rdkit} descriptors, (2) kNN on Morgan fingerprints (bit vector)~\cite{rogers2010extended}, (3) kNN on count-based Morgan fingerprints, and (4) kNN on Lorentz drug embedding without ATC information. We use the Tanimoto coefficient~\cite{bajusz2015tanimoto} as the similarity metric for fingerprints-based representations. Performance is evaluated using area under the receiver operating characteristic curve (AUROC) and area under the precision-recall curve (AUPRC). Figure~\ref{fig:repo_comp} shows that the LDE with ATC information outperforms other drug representations by a large margin. Averaging across different $k$ values, the LDE with ATC information surpasses Morgan fingerprints, the second best representation, by $12\%$ (AUROC) and $15.8\%$ (AUPRC). Compared to LDE without ATC information, incorporating drug hierarchy in the embedding achieves a large gain of $33.6\%$ (AUROC) and $48.8\%$ (AUPRC). LDE's competitive performance on discovering repositioning opportunities are likely driven by the drug-drug similarity that is encoded in the ATC hierarchy.

\section{Conclusion}
We introduced a method for learning a high-quality drug embedding that integrates chemical structures of drug and drug-like molecules with local similarity of drugs implied by a drug hierarchy. We leveraged the properties of the Lorentz model of hyperbolic space and developed a novel hyperbolic VAE method that simultaneously encodes similarity from chemical structures and from hierarchical relationships. We showed qualitatively that our embedding recapitulates the hierarchical relationships in the ATC hierarchy. We showed empirically that the embedding can be used for drug repositioning. 

\bibliographystyle{IEEEtran}
\bibliography{main.bib}

\end{document}